\documentclass[sigconf,balance=false]{acmart}
\usepackage{popets}
\usepackage{amsmath}
\usepackage{amsthm}
\usepackage{graphicx}
\usepackage{caption}
\usepackage{subcaption}
\usepackage{paralist}
\usepackage[ruled,linesnumbered,vlined]{algorithm2e}
\usepackage{color,colortbl}
\usepackage{url}
\usepackage{amsfonts}
\usepackage{booktabs}
\usepackage[shortlabels]{enumitem}

\newcommand*{\rom}[1]{\expandafter\@slowromancap\romannumeral #1@}

\newcommand\sys{VeraSel }

\setcopyright{popets}
\copyrightyear{YYYY}

\acmYear{YYYY}
\acmVolume{YYYY}
\acmNumber{X}
\acmDOI{XXXXXXX.XXXXXXX}
\acmISBN{}
\acmConference{}
\begin{document}

\title[VeraSel]{VeraSel: Verifiable Random Selection for Mixnets Construction}


\author{Xinshu Ma}
\affiliation{%
  \institution{University of Edinburgh}
  \streetaddress{}
  \city{}
  \country{}}
\email{x.ma@ed.ac.uk}

\author{Florentin Rochet}
\affiliation{%
  \institution{University of Namur}
  \streetaddress{}
  \city{}
  \country{}}
\email{florentin.rochet@unamur.be}

\author{Tariq Elahi}
\affiliation{%
  \institution{University of Edinburgh}
  \streetaddress{}
  \city{}
  \country{}}
\email{t.elahi@ed.ac.uk}


\renewcommand{\shortauthors}{Ma et al.}

\begin{abstract}
  The security and performance of Mixnets depends on the trustworthiness of the Mixnodes in the network. 
  The challenge is to limit the adversary's influence on which Mixnodes operate in the network. A trusted 
  party (such as the Mixnet operator) may ensure this, however, it is a single point of failure in the event of
  corruption or coercion. 
  Therefore, we study the problem of how to \textit{select a subset of Mixnodes} in a 
  distributed way for Mixnet construction. We present \sys, a scheme that enables Mixnodes to be 
  chosen according to their weights in a distributed, 
  unbiased, and verifiable fashion using Verifiable Random 
  Functions (VRFs). It is shown that \sys enables any party to 
  learn and verify which nodes has been selected based on the 
  commitments and proofs generated by each Mixnode with VRF.
\end{abstract}


\maketitle

\section{Introduction}
\label{sec:introduction}

Mixnets~\cite{chaum1981untraceable} are a fundamental type of anonymous communication network 
(ACN) 
that enables strong \textit{metadata} privacy protection on the Internet. At their inception, Mixnets were 
considered impractical for wide-scale deployment due to their 
inherent performance limitations. However, there has been a recent resurgent interest in Mixnets, 
and many recent designs with stronger security guarantees as well as improved
scalability have been proposed in both academia~\cite{van2015vuvuzela,
	kwon2017atom, kwonxrd, tyagi2017stadium, piotrowska2017loopix, chaum2017cmix,
	karaoke} and industry~\cite{diaz2021nym}.


Mixnets route traffic over multiple 
network relays (Mixnodes) that batch and/or delay messages to provide meta-data privacy, thus 
mitigating the threat of powerful adversaries linking communicating parties together. It is clear that the 
privacy of the communications depends on the trustworthiness of the Mixnodes. Indeed, the security of 
many known Mixnet-based designs~\cite{van2015vuvuzela,karaoke,piotrowska2017loopix,nym2021} 
relies on the \textit{anytrust} assumption where at least one Mixnode in the communication route must be 
honest. Communications over routes with only adversarial Mixnodes may be fully compromised, and 
compromise rates scale linearly with the resource (Mixnode) endowment of the adversary.

In practice, the \textit{anytrust} assumption is challenging to maintain for two related reasons. First, ACNs 
require resources to operate that may be beyond the capabilities of a single, or small group of, entity(ies) 
and which may also be a single point of failure. Second, users of an ACN do not necessarily trust the 
operator of the network, nor does the operator 
desire the attention and threat of coercion from powerful adversaries. It is thus safer to distribute the 
trust over non-colluding parties. To address both of these aspects, crowd-sourcing resources is a viable 
approach. 
A real-world example is Tor, which is 
operated by around $7000$ volunteer relays providing a total of almost $800$\,GB/s of bandwidth 
located in 
diverse jurisdictions around the world. 
In a similar fashion, 
The recent Nym 
Mixnet~\cite{diaz2021nym}, incentives third-parties to operate Mixnodes with cryptocurrency-based 
rewards. In both systems, an adversary may freely join the network by operating nodes and thus the 
challenge is to reduce the resulting privacy impact. 


In contrast to the Tor network, which aims to utilize \textit{all} available relays to maximize 
throughput, Mixnets must balance throughput and privacy simultaneously. This is because a 
Mixnet with too many Mixnodes creates the possibility that a communication traverses through the 
network without sufficient mixing---which requires many communications to transit the same Mixnode in 
a given time frame. Hence, a Mixnet may only require a subset of all available Mixnodes for secure 
operation. In order to 
obtain good performance, while mitigating against Sybil attacks, the Mixnet may select nodes based on 
their `weights', for example the bandwidth of the Mixnode. 
In other words, the Mixnet operator wants to select a fraction of all available Mixnodes $G$, in proportion 
to their 
weights, to form an active set $A$ that will be used in the construction of the Mixnet.

The challenge is to prevent the adversary from 
influencing the selection process and being disproportionately represented, with respect to their weight, 
in the selected Mixnode subset.
Therefore, the
selection process should be incorruptible even though there is no mutual trust between the Mixnodes or 
users 
of the Mixnet (the parties). One approach is to use a Trusted-Third Party (TTP), however, this 
introduces a single point of failure and potential collusion vector that could deviate from the selection 
process and ensure a disproportionate number of adversarial Mixnodes are selected into the active 
set.


To address this problem, we propose the \textit{verifiable random selection} (VeraSel) 
scheme that allows Mixnodes to be selected in a distributed, unbiased, and verifiable fashion. 
All parites in \sys are able to discover which Mixnodes have been selected into the active set, by 
processing the relay information, pseudo-random commitments, and proofs posted 
to the public bulletin board by Mixnodes.
Our contributions are mainly pragmatic rather than theoretical, building on existing cryptographic primitives to produce a distributed and verifiable random selection protocol.
To the best of our knowledge this is the first nodes selection protocol of its kind for the anonymous communication system.
\section{Background}
\label{sec:background}

In this section, we describe current Mixnet construction methods and 
the building blocks---VRF---used in our solution.

\subsection{Mixnets Construction}
\label{sec:mixnet construction}

\begin{figure}[h]
\centering
\includegraphics[width=.95\columnwidth]{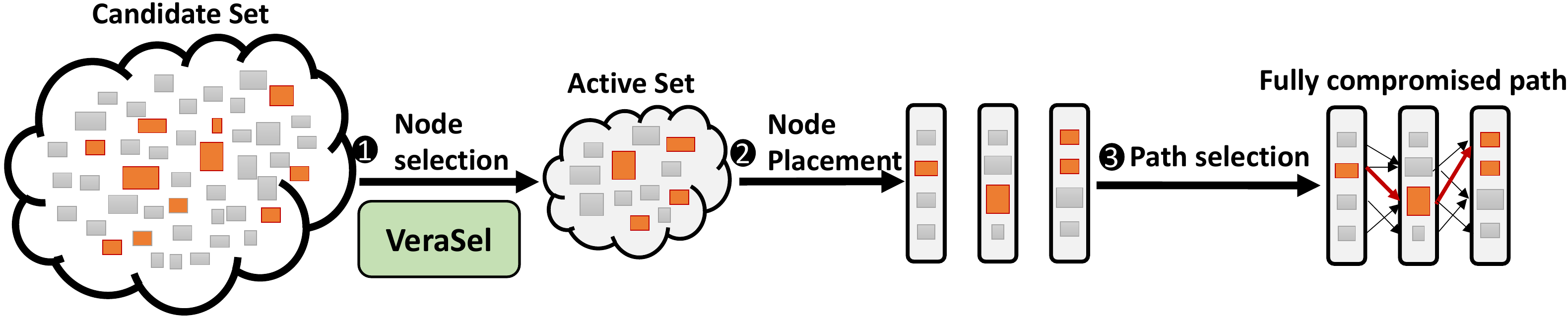}
\caption{Three general steps to construct a stratified Mixnet. Orange rectangles 
represent
malicious Mixnodes, while white objects are benign.}
\label{fig:model}
\end{figure}

Mixnets can be arranged in many topologies: mesh, 
cascade, and stratified are some of the most common. In this paper, we focus on the stratified topology~\cite{danezis2003mix} 
due to the evidence that it is both as, or more, secure and performant as the other two~\cite{dingledine2004synchronous, diaz2010impact}.
Real-world Mixnets are not static systems: they need to adapt to node churn and scale to network 
demand, and to accommodate a stratified network may be periodically (re)constructed from a random 
subset of the available Mixnodes.
Figure~\ref{fig:model} shows the general three steps of constructing a Mixnet 
where a subset of Mixnodes \textit{active set} are selected from \textit{candidate set} and each selected node will operate in the next epoch. For detailed illustration of why only using a subset of Mixnodes, please refer to Appendix~\ref{app:not_all}.

We briefly describe the Mixnet construction method in the recently released Nym~\cite{nym2021} system. 
To construct a network for the coming epoch, with a large number of available Mixnodes, a subset is 
selected (\textit{node selection}) to 
route traffic in Nym with a probability that is proportional to 
the nodes' stake. The selected Mixnodes are uniformly placed into layers in a 
distributed manner (\textit{node placement}). However, the specific design of selecting Mixnodes weighted by their stake is not discussed in detail~\cite{diaz2021nym}. Herein we identify the unsolved problem of incorruptible node selection in Mixnets construction
and propose \sys that is unbiased, verifiable, and applicable to the distributed system. 

\subsection{VRF}
\label{subsec:vrf}

Given an input $x$ (secret or not) and a private key $sk$, a
VRF~\cite{dodis2005verifiable}
returns $y$, a commitment to $x$ and a proof that $y$ commits to the value $x$.
We note that $(y, \pi) \leftarrow VRF(sk, x)$. A VRF is deterministic and
pseudorandom. That is, the commitment to $x$ is indistinguishable from a value
taken uniformly at random from the VRF's output space, and for two same input
pairs, the same output is derived. In our use-case, we require a pseudorandom
value in $\mathbb{Z}_{2^q}$ with $2^q$ greater or equal to the maximum possible
bandwidth in the network. We use a hashing function $H: \{0, 1\}^* \to \{0, 1\}^q$
to compute $H(y)$ under the uniform hashing assumption to capture the
requirement. Moreover, the pseudorandom value can only be computed by the
secret holder, cannot be predicted by any other party in the network and can be
verified by anyone in possession of the paired public key.

\section{\sys Design}
\label{sec:design}

\subsection{Protocol Overview}

\sys is an algorithm for selecting a weighted subset of Mixnodes 
in a distributed and verifiable 
fashion, without the dependence on a trusted third-party.
We formalize the node selection problem as follows: Given a set $G$ of Mixnodes, we want to sample a 
weighted subset of $G$, the \textit{active set} $A$ such that the sum of the
weights of the 
relays in the subset 
meets or just exceeds a predetermined fraction, $\tau$, of the total weight of the relays in $G$.


We assume the existence of PKI and that all Mixnodes can produce and post their public key material to a 
public bulletin board. We also assume that the Mixnet periodically refreshes its topology, both in 
terms of Mixnodes and connectivity between nodes, which requires a fresh active set every epoch.
In contrast to a centralized scheme where a trusted third party 
announces the active set for each epoch, clients in \sys are able to
securely produce the active 
set on their own, by using materials posted by the Mixnodes to a public bulletin board (see 
Appendix~\ref{subsec:bb} for a general description). 

\sys mobilizes VRFs, which provide deterministic and verifiable outputs that are pseudorandom and 
unpredictable. By using a public random seed value, \sys allows all Mixnodes to produce fresh 
randomness using their secret keys and the seed with the VRF. This fresh
randomness is a deterministic commitment to the seed and may then be used 
by any client to produce an intermediate data structure (see below) that can 
deterministically produce an verifiable active set that is common to all clients. 

In order to select Mixnodes according to their weight, $w$, \sys introduces the 
\textit{Weight Table} (WT). Initially, the size of $WT$, 
$W$, is the total weight of all the Mixnodes. Each Mixnode maps to an interval
in WT of the size of its weight, as shown in Figure~\ref{fig:weight table}.

Mixnodes are sorted according to their VRF output and inserted into the $WT$ in
that order. This sorting is stable assuming collisions on
the VRF output occurs with negligible probability, and produces the same sorted
list for all parties. To select nodes we again sort the Mixnodes 
according to their VRF output. We then take each Mixnode's random output modulo 
$W$ to find a uniformly random index into $WT$. The Mixnode that is mapped from that index is moved to 
the active set. This process is continued until the desired fraction of the total bandwidth is allocated to 
the active set. Through this uniform random selection from $WT$ we obtain a
weighted selection of Mixnodes that is identical for all parties.

\subsection{VeraSel Detail Description}
\label{subsec:detail}


%

In VeraSel, time is discretized into epochs and each epoch, $e$, is further partitioned into three 
non-overlapping time spans: \textit{post}, \textit{setup}, and \textit{select} in that order.


\begin{enumerate}
\item \textbf{Post.} 
A Mixnode, $i$, that wishes to join (or remain in) the network publishes its
public key, its weight, and a 
signature
$$p_i = \langle pk_i, w_i, Sign_{sk_i}(w_i)\rangle,$$ 
to 
the 
bulletin board. Note that  a Mixnode's public key acts as its unique
identifier.
A publication occurring during the \textit{post} part of the current epoch,
$e$, takes effect in the 
next epoch $e+1$. 

%

\item \textbf{Setup.} We assume that during the setup phase in each epoch a \textit{seed} is available 
  with the properties that it is pseudorandom, unpredictable, and global
(See discussion in 
Section~\ref{sec:random seed}). During Setup phase, a Mixnode $i \in G$ 
computes the VRF and publishes the outputs on the bulletin board as follows:
\begin{enumerate}[a)]
\item Generate a commitment and a \textit{proof} by computing 
  $$\langle y_i, \pi_i\rangle \leftarrow \text{VRF}(sk_i, seed),$$ 
where $sk_i$ is node's secret key.

\item Prepare and post the message on the bulletin board
  $$r_i = \langle y_i, \pi_i, Sign_{sk_i}(y_i||\pi_i) \rangle.$$
\end{enumerate}

\begin{figure}[!t]
		\centering
		\includegraphics[width=0.5\textwidth]{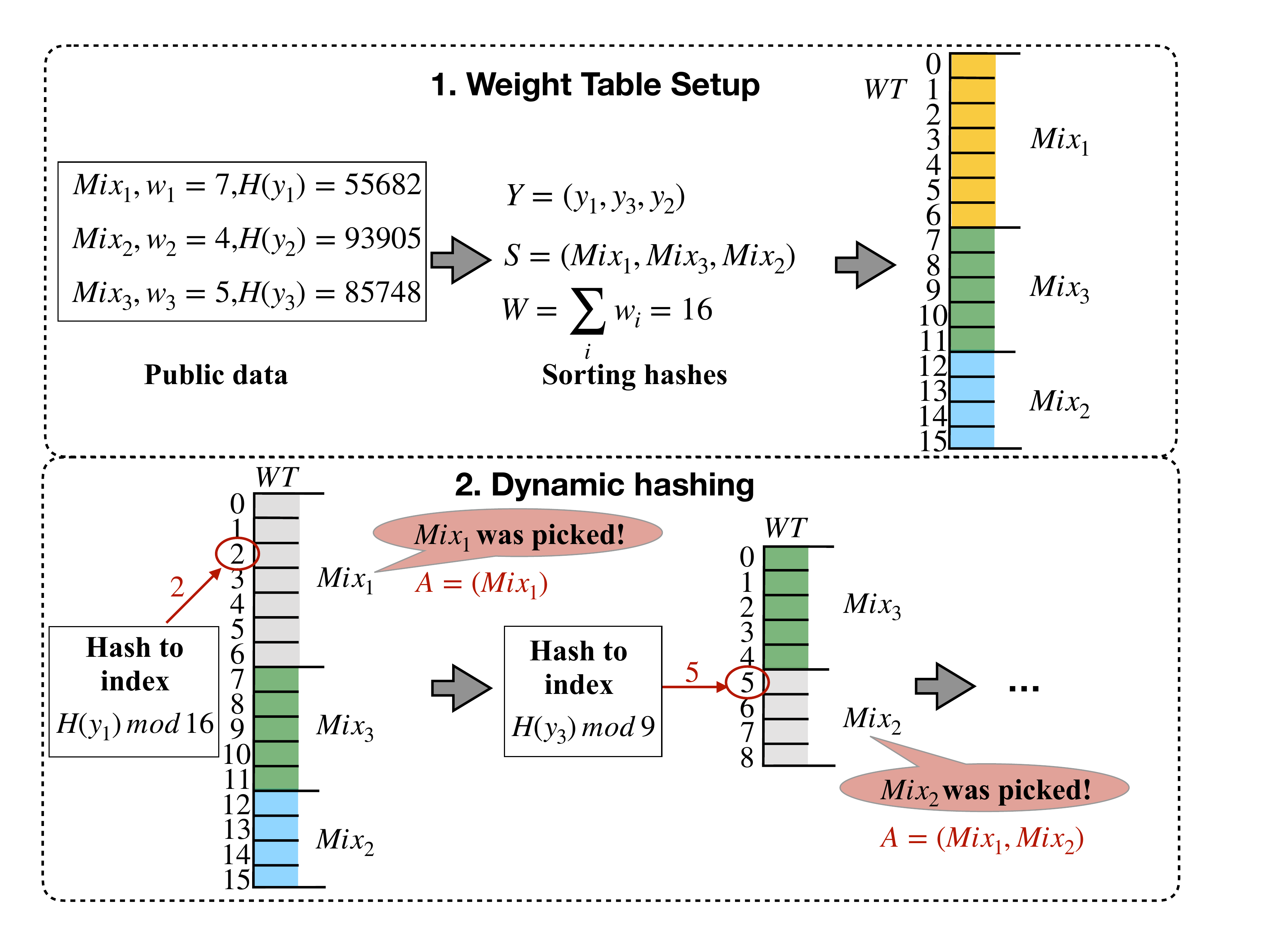}
		\caption{With the public data on the bulletin board, clients update with the latest topology by setting up a weight table (\textit{select a-c}) and performing dynamic hash (\textit{select d-f}) using public data on the bulletin board.}
		\label{fig:weight table}
\end{figure}

\item \textbf{Select.} 
The client downloads all $p_i$ and $r_i$ for the current epoch from the
bulletin board. It checks that the signatures match the expected identity,
and then proceeds (see Figure~\ref{fig:weight table}) as follows:


\begin{enumerate}[a)]
\item Check the correctness of each $y_i$
    $$\textit{VALID/INVALID} \leftarrow \text{VerifyVRF}(pk_i, y_i, \pi_i, seed),$$
and discard any Mixnodes with invalid commitments. 

\item Sort all valid Mixnodes in ascending order of their commitment values $H(y_i)$ and store in $Y = 
sort(y_i, y_k, ..., y_z)$.

\item For each Mixnode $i$ in $Y$ append $w_i$ intervals to $WT$.
The 
process 
concludes with $W=|WT|$.

\item Calculate 
  $$idx \leftarrow H(y_i)\;\text{mod}\;{W}.$$ 
\item Map $idx$ to the corresponding Mixnode $k$ that owns the interval $[begin,\;\; begin+w_k]$ such that 
$begin \leq idx \leq begin+w_k$.
Remove the selected node $k$, and update $W = W - |w_i|$.
\item Check if the total bandwidth of selected Mixnodes $A$ meets the threshold $\tau$. If yes, the selection 
stops and outputs the chosen subset $A$. Otherwise, go to step d).
\end{enumerate}

\end{enumerate}

\subsection{Discussion on \textit{Seed} Generation}
\label{sec:random seed}
For \sys, each epoch requires a \textit{seed} that is chosen at random and publicly known by every node. Indeed, the randomness of the selection procedure comes from this \emph{seed} and thus the adversary should not be able to manipulate it; otherwise, the adversary may choose a seed to obtain a VRF output $y$ that favors the selection of malicious nodes.
Thus,
\sys requires the \textit{seed} in every epoch to be unpredictable and unbiasable. Namely, the random \textit{seed} must be generated in such a way that no one can predict and knowingly bias the value to anyone's advantage or disadvantage. 


To generate a $seed$ for epoch $e$, in a similar manner to Algorand~\cite{gilad2017algorand}, the Mixnode $g$ with the minimum commitment value during last epoch $e-1$ computes a proposed $seed_e$ using VRF and the previous seed: 
$\langle seed_e, \pi_e\rangle \leftarrow \text{VRF}(sk_g, seed_{e-1}||e).$ The $sk_g$ is chosen well in advance of seed generation (Section~\ref{subsec:detail}), 
which ensures that $seed_e$ is pseudo-random even though $g$ is malicious. If $seed_e$ is invalid or $g$ does not generate the seed, the associated seed for epoch $e$ is computed as $seed_e \leftarrow H(seed_{e-1}||e)$. The $seed_0$ can be generated at the inception of Mixnet system by all initial participants using 
``commit-and-reveal'' approaches (see Appendix~\ref{app:randomgen}).




\subsection{\sys Extensions}
\sys solves the problem of periodically selecting a subset of Mixnodes that will actively operate in the network in a distributed way, and here we show that \sys could smoothly fit into a complete process of Mixnet construction (Section~\ref{sec:mixnet construction}).

To 
construct an $l$-layer stratified Mixnet from a set $G$ of candidate Mixnodes, the execution runs in 
two steps:

\begin{inparaenum}
\item Use \sys to select a subset of Mixnodes $A \leftarrow \text{VeraSel}(G)$, such that Mixnodes are randomly selected by their weights and the total capacity of $A$ is no less than $\tau$.

\item Each Mixnode $i \in A$ gets its layer index for the current 
epoch by computing $y_i\, \text{mod}\, l$. Since $y_i$ is the VRF 
commitment published on the bulletin board during the last step, any 
party can easily verify the position of Mixnodes by checking the correctness of $y_i$ (see Section~\ref{subsec:detail}.3a).
\end{inparaenum}

In this way, the Mixnet is constructed in a fully distributed 
manner with Mixnodes selected according to their weights and 
placed randomly into different layers (such as Nym's Mixnet). It 
is worth noting that \sys can also be easily altered to fit other 
Mixnet construction algorithms (e.g., Bow-Tie~\cite{ma2022stopping}) by providing the appropriate candidate set and weights as input.

In a broad sense, \sys itself is a generic tool 
that helps to select a weighted subset from a given set of candidate 
items in a distributed and verifiable way. In the future, we will investigate
other application scenarios of \sys in a decentralized system, such as cascade Mixnet construction.





\section{Evaluation}
\label{sec:evaluation}

\subsection{Security Analysis}
In this section, we show that \sys provides availability, unbiasability, and verifiability properties.
In the discussion below, we assume that each honest Mixnode follows the
protocol and the cryptographic primitives \sys uses provide their intended
security properties. Specifically, the security properties of VRF are
maintained even if the Mixnet operator adversarial.

\textbf{Availability.} Our goal is to ensure that all clients can successfully
complete the protocol and obtain the output, even in the presence of malicious
Mixnodes that behave arbitrarily. 
All information are published and consistent after \textit{post} and
\textit{setup} phase, which enables the consistency of the selected result.
This is true under the assumption that the VRF outputs are collision-free. Let the
output size be a pseudorandom value of $n$ bits, and $m$ Mixnodes commitments. A collision would happen with
probability $\frac{m(m-1)}{2^{n+1}}$, which is a negligible function of $n$.

A malicious Mixnode can always abort the protocol or simply not run it, such as
not updating their latest status or not committing the random \textit{seed},
leading to not being chosen to operate in the Mixnet. In effect, the
adversary achieves a ``self-DoS'' and, as long as there are available honest Mixnodes, the clients still 
reach the same subset $A$ and use the Mixnet.
Invalid postings and commitments submitted by the malicious mixnodes have the
same effect.

\textbf{Unbiasability.} We want to ensure that an adversary cannot influence
the final result $A$ of the protocol. A malicious Mixnode could bias the
selection result by manipulating the inputs of VRF such that the output $y$
favors himself. It is required that one input $sk$ should be fixed
(\textit{post} phase) before revealing another input---the random \textit{seed}
(\textit{setup} phase) of VRF---and thus it is extremely hard for the malicious
node to pick the right $sk$ without knowing the \textit{seed}. It would require
publishing a number of keys, say $k(n)$, such that the VRF output matches a given
value for at least one of the keys. This would happen with probability $\frac{k(n)}{2^n}$, which means
$k(n)$ is not poly-bounded, thus impossible to realize. Furthermore, the
unpredictability of VRF limits the adversary's ability to manipulate honest
Mixnodes' commitments, as it requires recovering other nodes' $sk$s. This
property directly translates to the value of $idx \in [0, W]$ (\textit{select}
phase)
that is derived from $y$ and used for the weighted selection; it is uniformly
distributed in this interval under the assumption of uniform hashing. 
By mapping each
Mixnode that participates in the current epoch to a non-overlapping
interval in $[0, W]$ proportional to its bandwidth, a given computed
$idx$ value always falls into one interval and selects the related node. 

\textbf{Verifiability.} We want to ensure that the final result $A$ is verifiable. In \sys, every party that is 
able to access the public bulletin board can obtain the final selection result $A$. The bulletin board 
contains all information about the Mixnode weights, commitments, and proofs. Any verifying party can 
decide on its validity since all data are singed or committed with secret key and both signatures and 
commitments are verified using the corresponding public keys.
If the verifying party finds these data acceptable, he can replay the \textit{select} execution and verify 
whether or not any particular Mixnode is selected.
After a successful protocol run completes, any party can independently verify $A$.

\subsection{Performance Analysis}
\label{sec:performance}
This section evaluates our implementation of VeraSel. The primary question we wish to answer is 
whether \sys provides a statistically correct selection result. The second important question is 
what the communication and computation costs are.
\subsubsection{Correctness Validation}
We implemented our \sys protocol on the popular Curve25519 elliptic 
curve with Elliptic Curve VRF~\cite{rfc2020vrf}; we also implemented 
a simple \textit{unverifiable} weighted selection by a trusted party as the evaluation 
benchmark. We simulated the scenario where we want to select a subset of nodes $A$, weighted by their 
bandwidth, from a pool of $1000$ candidate Mixnodes $G$
such that the total bandwidth of $A$ is $\tau = 50\%$ of $G$. In the 
experiment setting, the total bandwidth of the $G$ set is $W\approx9970\,MBps$; the desired bandwidth 
of $A$ is $\approx 4985\,MBps$. 
The bandwidths of Mixnodes are generated by fitting to the bandwidth 
distribution of Tor relays from its historical data~\cite{CollecTor}.

\begin{figure}[!h]
		\centering
		\includegraphics[width=0.25\textwidth]{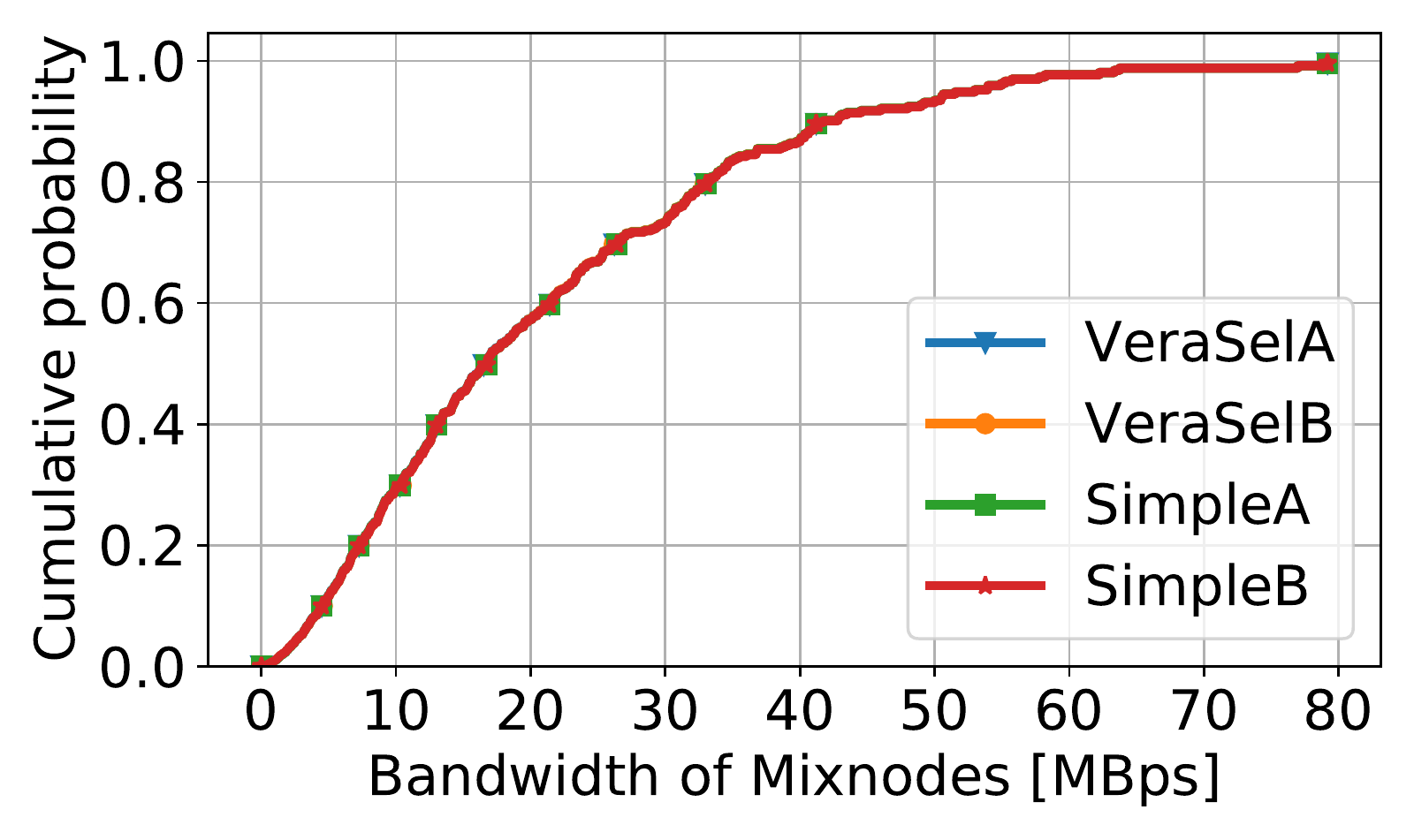}
		\caption{Probability distribution on selecting Mixnodes with VeraSel and Simple selection. }
		\label{fig:cdf}
\end{figure}

We run \sys and Simple selection on the same set of Mixnodes; the results are derived from $1000$ runs (test ``A'') and $3000$ runs (test ``B'')respectively shown in Figure~\ref{fig:cdf}.
Overall, we can see that in all tests the expected selection rates for the same Mixnodes with \sys are 
almost the same with Simple selection. Furthermore, the results of Kolmogorov-Smirnov test confirm this 
conclusion: at $\alpha = 0.05$ and both sample sizes are $3000$, $D_{\alpha}=0.0351$ and 
$statistic=0.01$.

\subsubsection{Communication and Computation Cost}
Suppose there are $n$ Mixnodes in set $G$, \sys has linear communication complexity and uses only 
three rounds. Although the dominant cost is VRF computation, the \sys protocol is fast to 
execute with a subsecond computation cost per Mixnode and client.


\section{Conclusion}
\label{sec:conclusion}
In this paper, we address the problem of \textit{how to select a subset of Mixnodes in a distributed way} 
for Mixnet construction. We present \sys, an algorithm that enables Mixnodes to be selected according to 
their weights in a distributed, unbiased, and verifiable fashion. \sys enables any party to learn and verify 
which nodes have been selected based on the commitments and proofs generated by each Mixnode with 
VRF. Moreover, we validated the correctness of \sys and the empirical simulation shows it to be practical.


\bibliographystyle{ACM-Reference-Format}
\bibliography{ref}

\appendix
\section{Argument for using a subset of Mixnodes}
\label{app:not_all}
\begin{figure}[h]
  \centering
  \begin{subfigure}{.25\textwidth}
    \centering
    \includegraphics[width=.91\linewidth]{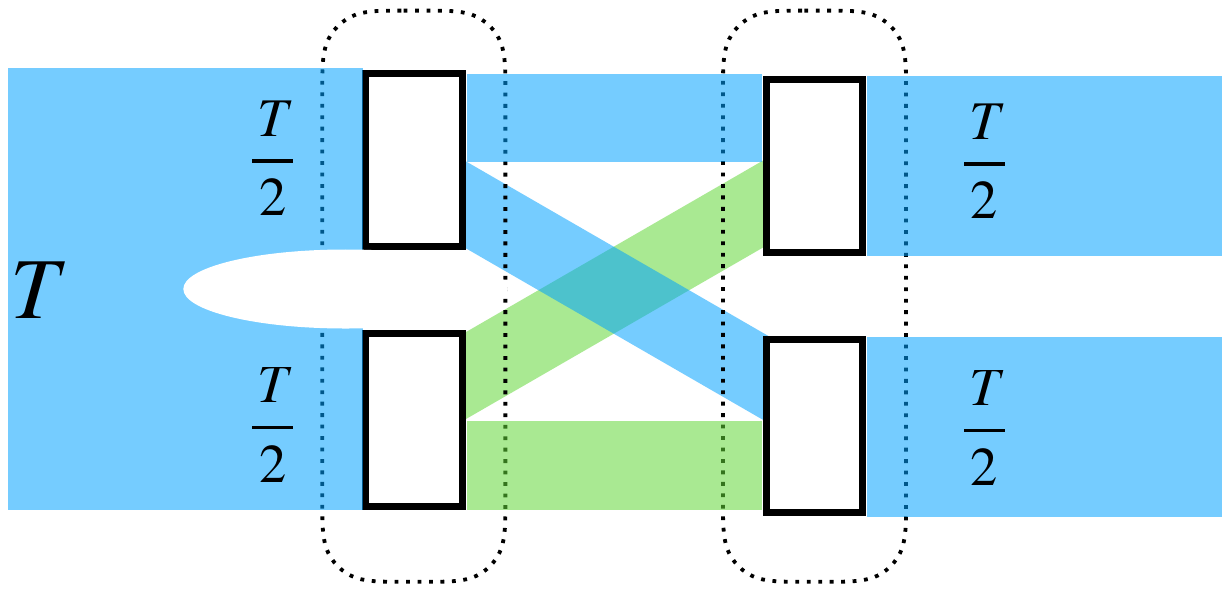}
    \caption{A $2 \times 2$ stratified mixnet.}
    \label{fig:mix22}
  \end{subfigure}%
  \begin{subfigure}{.25\textwidth}
    \centering
    \includegraphics[width=.84\linewidth]{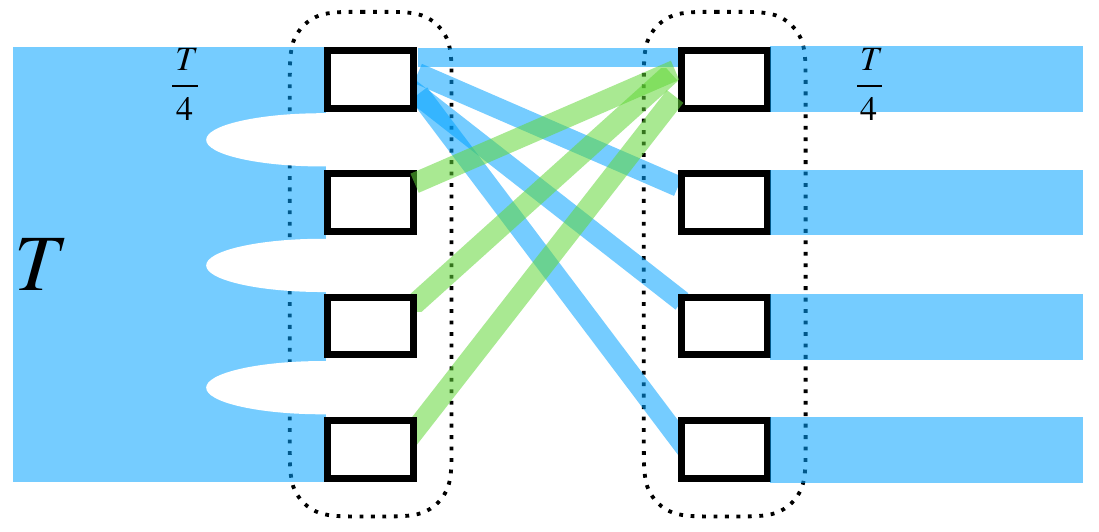}
    \caption{A $4 \times 4$ stratified mixnet.}
    \label{fig:mix42}
  \end{subfigure}
  \caption{Comparison between two $2-$layer stratified Mixnets with different number of MixNodes.}
  \label{fig:thick_thin_traffic}
\end{figure}

There are security implications of
involving a large number of relays in Mixnets, in contrast to Tor where almost
all relays are utilized for performance reasons. More specifically, 
from the perspective of an \textit{external global passive 
adversary}, as the
number of mix nodes per layer increases there is more spreading of the incoming
traffic and relatively less traffic at individual MixNodes 
(i.e., a ``diffusion'' of
traffic), which intuitively would decrease the probability of mixing of messages and lower the system 
entropy, as 
depicted in Figure~\ref{fig:thick_thin_traffic}. 

In Figure~\ref{fig:mix22} we see that the total traffic is split across two 
mixes in a ``concentrated'' traffic per mix in a $2\times 2$ 
stratified mixnet. Contrast this to the more ``diffuse'' traffic 
per server in a $4\times4$ stratified mixnet in Figure~\ref{fig:mix42} where the same 
traffic is split in four. Intuitively, the entropy of the mix 
network in Figure~\ref{fig:mix42} will be lower as compared to that in Figure~\ref{fig:mix22} since fewer 
messages will arrive at any mix in the former, and hence be mixed 
with fewer messages over all.

\section{Bulletin Board} \label{subsec:bb}

We require a consistent and secure view to all involved 
parties in a Mixnet.
This can be realized by a publicly accessibly and reliable bulletin 
board which allows users to post and read all necessary 
information. Our system requires the bulletin board to have the 
following properties: \begin{inparaenum}\item only data that have 
been posted to the bulletin board appear on it; \item data cannot 
be modified or erased from the bulletin board once it is published.
\end{inparaenum}

Although implementing a public bulletin board is beyond our scope, we note that it can be achieved in a Byzantine standard threat model (i.e., honest nodes are the majority). In the literature, depending on the implementation, the bulletin board could be centralized~\cite{fujioka1992practical,peters2005secure,lenstra2017trustworthy} or distributed~\cite{kiayias2018security,culnane2014peered}. Here, to simplify the problem, we assume the bulletin board service is provided by a trustworthy server.

\section{Randomness Generation}
\label{app:randomgen}
Generating a random number is a classical problem and has been well studied in the 
literature. One traditional approach is to rely on a trusted third 
party to emit randomly chosen integers regularly to the public (such 
as Rabin's ideal random beacon~\cite{rabin1983transaction}, NIST beacon~\cite{nistbeacon}). 
To get around the centralized beacon, one approach is to extract 
random values from a public entropy source such as stock-market data~\cite{clark2010use}, 
lottery~\cite{baigneres2015trap}, or Bitcoin~\cite{bonneau2015bitcoin}. However, while each of these 
data 
sources are 
hard to predict, they are vulnerable to the malicious insiders (e.g., 
high-frequency traders, lottery administrators, Bitcoin miners).
An alternative approach is to generate random outputs jointly by a 
collection of participants (i.e., nodes) and the common paradigm is 
``commit-and-reveal'' where each party $i$ commits to a 
nonce $r_i$ before all commitments are revealed and the random 
output is computed as $r = \oplus_i r_i$. Many protocols based on 
Publicly Verifiable Secret Sharing (PVSS) (~\cite{schindler2020hydrand,syta2017scalable,kokoris2020asynchronous,cascudo2020albatross,cascudo2017scrape}) 
where each node uses PVSS to share the nonce requires super-linear communication and multi-round interactions. 
Other protocols based on Verifiable Random Functions (VRF) (~\cite{gilad2017algorand,david2018ouroboros,kiayias2017ouroboros}) 
where a leader generates the random output with VRF requires a leader sortition scheme. 
Both of them are able to at best resist an 
adversary that controls minority nodes (i.e., less than $50\%$ 
malicious resource) while the protocols based on Verifiable Delay Function (VDF)~\cite{ephraim2020continuous} or sloth~\cite{lenstra2015random} have linear communication complexity and produce unpredictable randomness even with $n-1$ malicious parties.

\end{document}